\documentclass[aps,prl,twocolumn,amsmath,amssymb,footinbib,showpacs,longbibliography,superscriptaddress]
{revtex4-1}

\usepackage[english]{babel}
\usepackage{latexsym}
\usepackage{graphics}
\usepackage{subfigure}
\usepackage{epsfig}
\usepackage{color}
\usepackage{hyperref}
\usepackage{bbold}

\usepackage{braket} 

\hypersetup{
colorlinks=true,
citecolor=blue,
linkcolor=red,
urlcolor=black
}

\newcommand{\epsfboxmod}[1]{\epsfbox{#1}}

\newcommand{\infig}[2]{\begin{center}
                                    \mbox{ \epsfxsize #1 \epsfboxmod{#2}}
                                    \end{center}}

\newcommand{\ie}{\textit{i.e.}}

\newcommand{\betadens}{{\beta_{\textrm{\tiny dens}}}}

\newcommand{\betatr}{{\beta_{\textrm{\tiny tr}}}}

\newcommand{\Vr}{V_{\textrm{\tiny R}}}
\newcommand{\Ur}{U_{\textrm{\tiny R}}}
\newcommand{\sigmar}{\sigma_{\textrm{\tiny R}}}

\DeclareMathOperator{\dd}{\mathrm{d}\!}

\DeclareMathOperator{\e}{e}

\renewcommand{\section}[1]{\textit{#1.---}}

\begin{document}

\title{Effect of a bias field on disordered waveguides: Universal scaling of conductance and application to ultracold atoms}

\author{C\'ecile Crosnier de Bellaistre}
\affiliation{
 Laboratoire Charles Fabry,
 Institut d'Optique, CNRS, Univ Paris-Saclay,
 2 Avenue Augustin Fresnel,
 F-91127 Palaiseau Cedex, France
}

\author{Alain Aspect}
\affiliation{
 Laboratoire Charles Fabry,
 Institut d'Optique, CNRS, Univ Paris-Saclay,
 2 Avenue Augustin Fresnel,
 F-91127 Palaiseau Cedex, France
}

\author{Antoine Georges}
\affiliation{
 Coll\`ege de France, 11 Place Marcelin Berthelot, F-75005 Paris, France
}
\affiliation{
 Centre de Physique Th\'eorique, Ecole Polytechnique, CNRS, Univ Paris-Saclay, F-91128 Palaiseau, France
}
\affiliation{
 Department of Quantum Matter Physics, University of Geneva, 24 Quai Ernest-Ansermet, CH-1211 Geneva 4, Switzerland
}

\author{Laurent Sanchez-Palencia}
\affiliation{
 Laboratoire Charles Fabry,
 Institut d'Optique, CNRS, Univ Paris-Saclay,
 2 Avenue Augustin Fresnel,
 F-91127 Palaiseau Cedex, France
}
\affiliation{
 Centre de Physique Th\'eorique, Ecole Polytechnique, CNRS, Univ Paris-Saclay, F-91128 Palaiseau, France
}

\date{\today}

\begin{abstract}
We study the transmission of a disordered waveguide subjected to a finite bias field.
The statistical distribution of transmission is analytically shown to take a universal form. 
It depends on a single parameter, the system length expressed in a rescaled metrics, which encapsulates all the microscopic features of the medium and the bias field. Excellent agreement with numerics is found for various models of disorder and bias field. For white-noise disorder and a linear bias field, we demonstrate the algebraic nature of the decay of the transmission with distance, irrespective of the value of the bias field.
It contrasts with the expansion of a wave packet, which features a delocalization transition for large bias field. The difference is attributed to the different boundary conditions for the transmission and expansion schemes.
The observability of these effects in conductance measurements for electrons or ultracold atoms
is discussed,
taking into account key features, such as finite-range disorder correlations, nonlinear bias fields, and finite temperatures.
\end{abstract}

\maketitle

\section{Introduction}
Anderson localization in \textit{unbiased} disordered materials is signaled by the
exponential suppression of diffusion and conductance~\cite{anderson_absence_1958,lee_disordered_1985,abrahams_50_2010}.
The connection between the two is firmly established by linear-response theory and the Einstein-Sutherland  relation~\cite{rammer_quantum_1998,akkermans_mesoscopic_2007}.
Hence, localized wave packets and transmission coefficients are characterized by the same exponential decay with distance.
Bias fields
induce a strong nonlinear response, 
which significantly affects localization and questions this relation. 
For a weak bias field, algebraic (rather than exponential) localization
has been established in previous 
numerical~\cite{soukoulis_localization_1983,bentosela1985} and analytical~\cite{prigodin_one-dimensional_1980,delyon_power-localized_1984} work. 
More precisely, Ref.~\cite{delyon_power-localized_1984} presented 
a rigorous proof that the eigenstates become extended beyond a critical value of
a dimensionless parameter $\alpha$, which characterizes the ratio of the bias force (opposite gradient of the bias field) to the disorder (see the precise definition of $\alpha$ below).
It is qualitatively consistent with a diagrammatic calculation of the asymptotic density of an expanding 
wave packet~\cite{prigodin_one-dimensional_1980}, yielding the average density $\overline{n(x)} \sim 1/x^{\betadens}$ in the direction of the bias force, 
with $\betadens=1+(1-\alpha)^2/8\alpha$ for $\alpha<1$ and where the overline denotes disorder averaging.
For $\alpha >1$, the asymptotic density is not normalizable, 
hence signaling a delocalization transition at $\alpha=1$.
In contrast, numerical evidence was provided in Ref.~\cite{soukoulis_localization_1983} that the transmission coefficient for the Kronig-Penney lattice model decays algebraically for arbitrary large bias force, yielding $\exp{\left(\overline{\ln T}\right)} \sim 1/x^\betatr$, and is thus unaffected by the delocalization transition.
Moreover, the exponent $\betatr \simeq 1/2\alpha$
is not trivially related to
the exponent $\betadens$.
These apparently contradicting behaviors raise several questions.
First, it is unclear whether the result of Ref.~\cite{soukoulis_localization_1983} is universal or model dependent.
In particular, the behavior of the characteristic quantity $\exp{\left(\overline{\ln T}\right)}$ in continuous-space models as considered in Ref.~\cite{prigodin_one-dimensional_1980} is unknown.
Second, available numerics do not provide the behavior of the average transmission $\overline{T}$, which would be more directly comparable to the average density $\overline{n}$.
Third, the behavior of physical quantities, in particular those directly related to the average transmission, such as Landauer conductances, remains unclear.
These questions have applications in mesoscopic physics, including the propagation of microwaves in inhomogeneous disorder~\cite{xu_light_2017} and the electric response to a bias field in disordered carbon nanotubes~\cite{seunghun2007} or silicon nanowires~\cite{hasan2013}, for instance.
It also applies to ultracold atoms where Anderson localization can be studied quantitatively~\cite{
sanchez-palencia_anderson_2007,shapiro_expansion_2007,kuhn_coherent_2007,
piraud2011,
billy_direct_2008,roati_anderson_2008,
kondov_three-dimensional_2011,
jendrzejewski_three-dimensional_2012,
semeghini_measurement_2015,
chabe_experimental_2008,
lopez_experimental_2012} (for reviews, see also Refs.~\cite{modugno2010,lsp2010}). 
In those systems, a bias field can easily be turned on, and
both conductance~\cite{brantut_conduction_2012,stadler_observing_2012,krinner_superfluidity_2013} and expansion dynamics~\cite{billy_direct_2008,roati_anderson_2008,kondov_three-dimensional_2011,jendrzejewski_three-dimensional_2012} are accessible.

Here, we study the transmission of a continuous one-dimensional disordered
waveguide subjected to arbitrary disorder and bias field.
The statistical distribution of transmission
is written in a universal form.
It is characterized by a unique parameter, the length of the waveguide expressed in a rescaled metrics, 
which encapsulates all microscopic features of the medium and bias field.
For white-noise disorder and a uniform bias force, we
derive analytically the relations $\exp(\overline{\ln T}) \sim 1/x^{1/2\alpha}$ and $\overline{T} \sim 1/x^{1/8\alpha}$
for arbitrary strength of the bias force.
We also perform numerical calculations for various models, and obtain excellent agreement with the
analytical results.
The different power law obtained for $\overline{T}$ in comparison to that of $\overline{n}$ found in Ref.~\cite{prigodin_one-dimensional_1980} is attributed to the different boundary conditions, whose role is enhanced by the long-range algebraic tails of the eigenstates
in the presence of a bias field.
Application to conductance measurements is discussed.
In particular, we include important realistic features, which significantly affect the conductance as measured in electronic or ultracold atomic systems, such as finite disorder correlations, nonuniform bias forces, and finite temperatures.

\section{Statistical distribution of transmission}
\begin{figure}[t!]
\infig{24em}{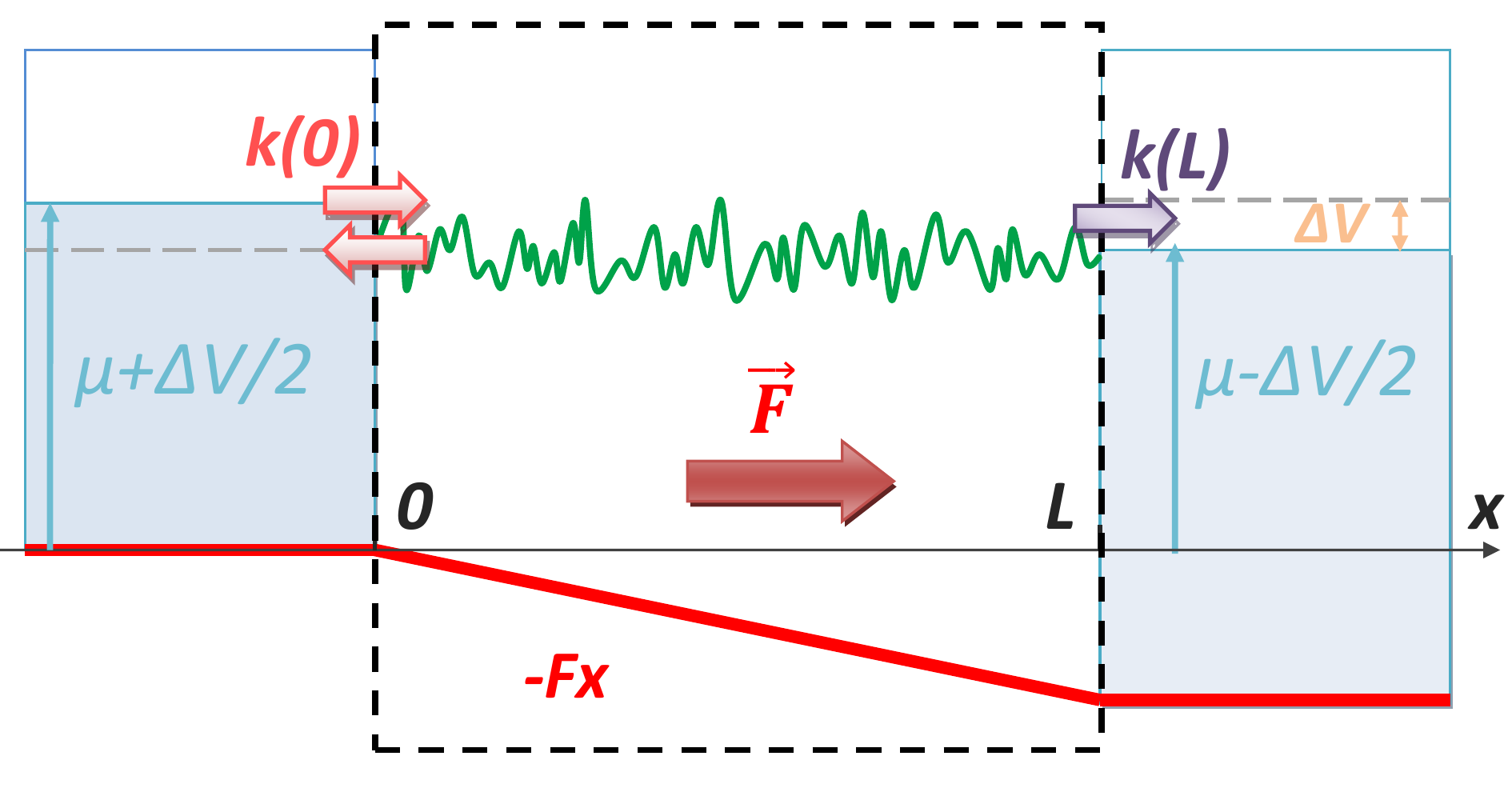}
\caption{\label{fig:reservoirs_deltaV}
Transmission and conductance of a matter waveguide of length $L$ (black rectangle)
in the presence of a bias force $F(x)$ and disorder (random green line).
The bias field is shown for a constant force (red line).
The incident and reflected wave vectors are $k(0)$ and the transmitted one is $k(L)$.
The Landauer conductance is measured from the discharge between two reservoirs (left and right blue boxes) with average chemical potential $\mu$ and infinitesimal potential difference $\Delta V$.
}
\end{figure}

To start with, consider the transmission of a coherent wave in a disordered material of length $L$ in the presence of a bias force $F(x)$ (see the dashed black rectangle in Fig.~\ref{fig:reservoirs_deltaV}).
In the following, we adopt the language of quantum matter waves for concreteness. The results are, however, directly applicable to other types of waves, such as linearly polarized microwaves or sound waves since the stationary states of all are governed by similar second-order space differential equations.
The disordered potential $V(x)$ is homogeneous and Gaussian.
Its average is zero and its two-point correlation function reads $C(x)\equiv\overline{V(x'+x)V(x')}$.
The latter may model a theoretical white-noise disorder, with $C(x)=\Ur\delta(x)$ and $\Ur$ the disorder strength,
or a more realistic correlated disorder, where $C(x)$ is some function decaying on the typical length scale $\sigmar$.

To compute the statistical distribution $P(T,x)$ of the transmission coefficient $T$ at the distance $x$,
we use the transfer-matrix approach, here generalized
to include a, possibly inhomogeneous, bias field.
In brief (see details of the derivation in the Supplemental Material), $P(T,x)$ is governed by the Fokker-Planck equation
\begin{equation}\label{eq:FPE}
\ell_-(x) \frac{\partial P }{\partial x} = \frac{\partial\, T^2 P}{\partial T}+\frac{\partial^2}{\partial T^2}\Big(T^2 (1-T)P\Big),
\end{equation}
with $\ell_-(x) \simeq {2\hbar^2K(x)}/{m\tilde{C}[2k(x)]}$,
$m$ the particle mass,
$E \geq 0$ the energy,
$K(x) \equiv E + \int_0^x\dd x'\, F(x')$,
and the initial condition $P(T,x=0)=\delta(T-1)$.
Equation~(\ref{eq:FPE}) has a straightforward physical interpretation.
We find the same equation as for unbiased disordered systems, except that
the backscattering mean free path $\ell_-(x)$ must be computed at the effective semiclassical kinetic energy $K(x)$
and is thus position dependent.
The validity of Eq.~(\ref{eq:FPE}) relies on the sole assumptions
that the disorder is weak, \ie, $\ell_-(x) \gg 1/k(x),\sigmar$ with $k(x)=\sqrt{2mK(x)}/\hbar$ the local wave vector,
and that the work of the bias force is negligible on the disorder correlation length, \ie,
$F(x)\sigmar \ll K(x), \hbar^2 k(x)/2m\partial_k\ln\tilde{C}[2k(x)]$.
Note that for a positive bias force $F$ and a bounded disorder power spectrum $\tilde{C}$, both the backscattering mean free path $\ell_-(x)$ and the wave vector $k(x)$ increase with the distance $x$. Hence the validity conditions of Eq.~(\ref{eq:FPE}) are always fulfilled, at least in the asymptotic limit $x \rightarrow \infty$.

The quantity $\ell_-(x)$ provides the natural metrics in the disordered material in the presence of the bias field, and, using the inhomogeneous dimensionless coordinate
\begin{equation}\label{eq:metrics}
s(x) = \int_0^x \frac{\dd x'}{\ell_-(x')},
\end{equation}
it disappears from Eq.~(\ref{eq:FPE}).
The same rescaling holds for inhomogeneous disorder (see, for instance, Ref.~\cite{xu_light_2017}).
Equation~(\ref{eq:FPE}) thus admits the analytic solution
\begin{equation}\label{eq:DistrAbrikosov}
P(T,L)=\frac{2\e^{-s(L)/4}}{\sqrt{\pi}s(L)^{3/2}T^2}\int_{\cosh^{-1}\sqrt{1/T}}^{\infty} \dd y \frac{y\ \e^{-y^2/s(L)}}{\sqrt{\cosh^2 \! y \! - \! 1/T}},
\end{equation}
\ie, the same as for unbiased, homogeneous disorder with $L/\ell_-$ rescaled to $s(L)$ (see, for instance, Ref.~\cite{abrikosov_paradox_1981}).
Note that in this universal form
all the microscopic features of the medium, such as disorder correlations and bias field, are fully encapsulated into the definition of the metrics~(\ref{eq:metrics}).
Formula~(\ref{eq:DistrAbrikosov}) is rigorous and allows us to compute exactly the various disorder-averaged quantities relevant to different questions.


\begin{figure*}[t]
\includegraphics[width=0.491\textwidth]{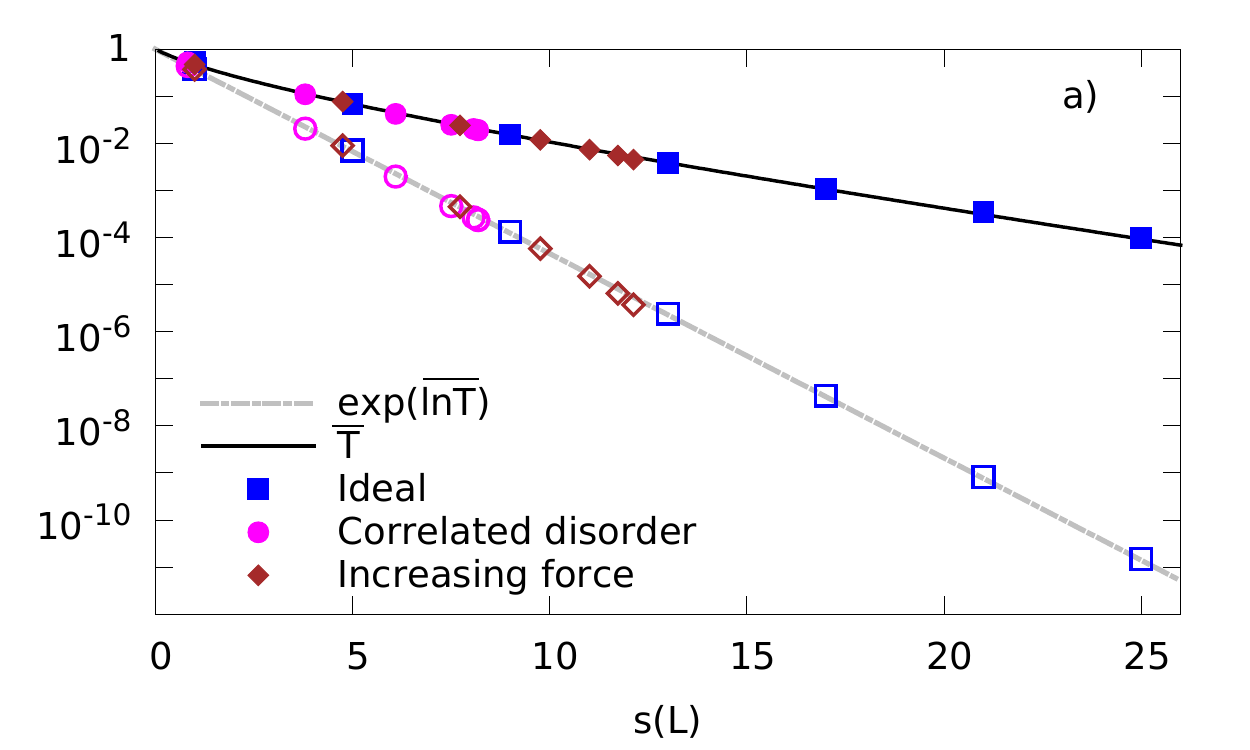}
\includegraphics[width=0.491\textwidth]{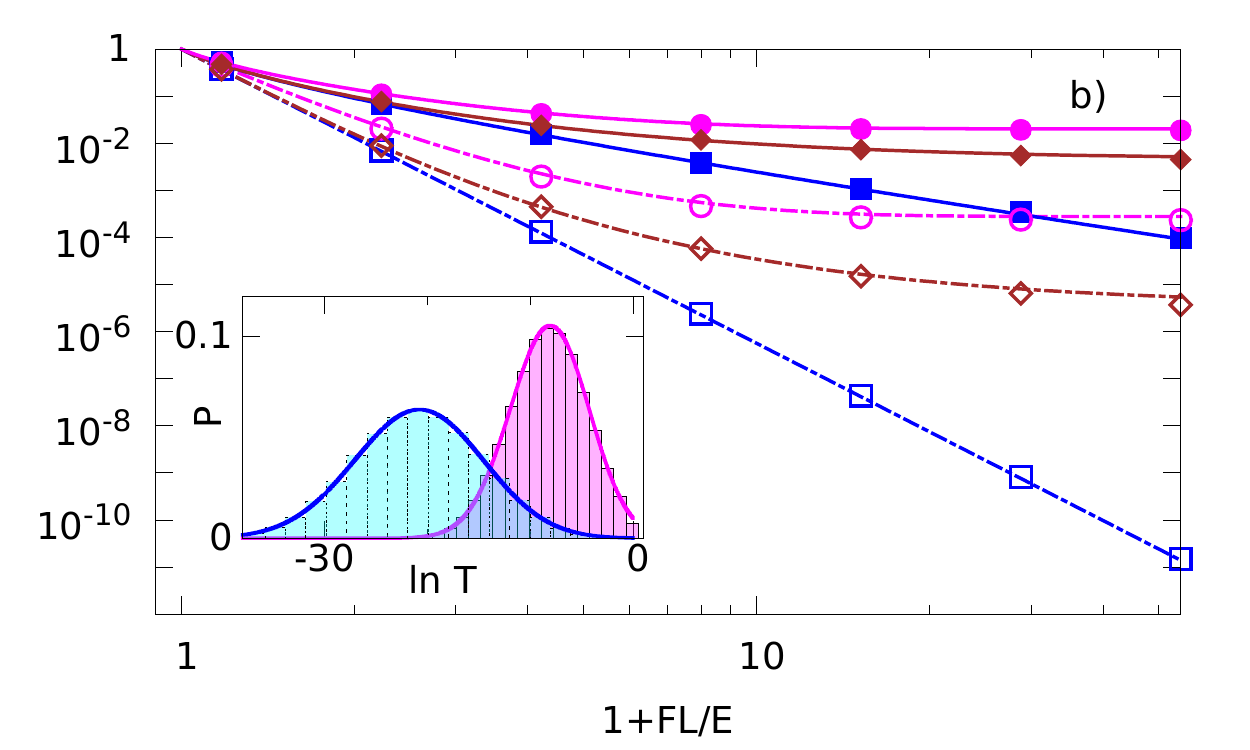}
\caption{\label{fig:graph_universality}
Analytical prediction vs numerical results for $\overline{T}$ (solid lines and solid symbols) and $\exp(\overline{\ln T})$ (dashed-dotted lines and open symbols),
for $\alpha=0.08$ and $2mF/\hbar^2 k^3(0)=0.01$
in the following cases:
uniform bias force and white-noise disorder (blue squares),
uniform bias force and Gaussian correlated disorder with $\sigmar k(0)=0.3$ (magenta circles),
harmonic bias force with $2m\omega/\hbar k(0)^2=0.01$ and uncorrelated disorder (brown diamonds).
(a)~Results plotted as a function of the metrics $s(L)$ [for white noise and constant force, $s(L) \propto \ln(1+FL/E)$; see Eq.~(\ref{eq:Swncf})].
(b)~Results plotted as a function of $1+FL/E$. 
Inset: Probability distribution of $\ln T$ for $FL/E=28$, as found from numerical simulations (histogram) compared to analytical prediction (line),
for a constant force, with correlated (dark, magenta) and uncorrelated (light, blue) disorder.
}
\end{figure*}

\section{Algebraic localization}
Let us start with the logarithm of the transmission.
Due to its self-averaging character, it represents the typical transmission and it is the quantity that is usually computed numerically, as in Ref.~\cite{soukoulis_localization_1983}, for instance.
Using Eq.~(\ref{eq:DistrAbrikosov}), we find $\overline{\ln T(L)} = - s(L)$.
This formula, together with Eq.~(\ref{eq:metrics}), exactly matches the heuristic formula proposed in Ref.~\cite{soukoulis_localization_1983} to interpret numerical results for the specific Kronig-Penney model.
Our analysis justifies this formula on rigorous grounds and generalizes it to any model of disorder and bias field.
For white-noise disorder and uniform bias force,
where $\tilde{C}(2k)=\Ur$ and $K(x)=E+Fx$,
we find 
\begin{equation}\label{eq:Swncf}
s(L)=\int_0^L \dd x \ \frac{m\Ur}{2\hbar^2(E+Fx)}=\frac{1}{2\alpha}\ln\left(1+\frac{F L}{E}\right),
\end{equation} 
where $\alpha={\hbar^2F}/{m\Ur}$ is the relative strength of the bias force and the disorder~\cite{prigodin_one-dimensional_1980}. It yields the characteristic algebraic decay $\exp\left(\overline{ \ln T}\right) \sim 1/L^{1/2\alpha}$. 
Excellent agreement with exact numerical calculations is found for continuous white-noise disorder [see the open blue squares and dotted-dashed line on Fig.~\ref{fig:graph_universality}(a)].

In order to compare the behavior of the transmission to the result of Ref.~\cite{prigodin_one-dimensional_1980} for the expanding wave packet, we now compute the average transmission. This quantity is also the one that determines various physically relevant quantities, such as the Landauer conductance (see below). 
Using Eq.~(\ref{eq:DistrAbrikosov}), we find the exact formula
\begin{equation}\label{eq:avT}
\overline{T(L)}=\frac{4\e^{-s(L)/4}}{\sqrt{\pi}s(L)^{3/2}}\int_0^{\infty}\dd y\ \frac{y^2\e^{-y^2/s(L)}}{\cosh(y)}.
\end{equation}
Again, it is in excellent agreement with exact numerical calculations [see the solid blue squares and the solid line on Fig.~\ref{fig:graph_universality}(a)].
For white-noise disorder and uniform bias force, we find the asymptotic
behavior $\overline{T(L)} \sim 1/L^{1/8\alpha}$, up to logarithmic corrections.
The difference of the scalings $\exp\left(\overline{\ln T(L)}\right) \sim 1/L^{1/2\alpha}$ and $\overline{T(L)} \sim 1/L^{1/8\alpha}$ originates from the well known large fluctuations associated with the statistical distribution~(\ref{eq:DistrAbrikosov})~\cite{beenakker_randum_1997}.

In turn, it is remarkable that the scaling $\overline{T(L)} \sim 1/L^{1/8\alpha}$ differs from that found for the density profile of an expanding wave packet, $\overline{n(x)} \simeq 1/x^{1+(1-\alpha)^2/8\alpha}$~\cite{prigodin_one-dimensional_1980}.
To make a direct comparison, we have computed the quantity $\overline{T(L)}$ using a different formalism than the transfer-matrix approach, namely the diagrammatic approach used in Ref.~\cite{prigodin_one-dimensional_1980} to compute the quantity $\overline{n(x)}$ in the expansion scheme. The latter needs to be adapted to the transmission scheme we consider here. Indeed, all diagrams involving scattering outside the region $[0,L]$ must be excluded. Taking this difference into account, the diagrammatic method allows us to recover the behavior predicted by the transfer-matrix approach. Physically, the strong difference between expansion and transmission can be understood as follows. Consider
a particle initially at position $x=0$ and look at the probability that it has been transmitted beyond $x=L$ after infinite time.
If the disorder is restricted to the space interval $[0,L]$, this probability is given by the transmission coefficient $\overline{T(L)}$ and therefore decays as $L^{-1/8\alpha}$.
In contrast, if the disorder extends over the full $x$ line, it turns into $\int_L^{\infty}\dd x\; \overline{n(x)}$, which decays as $L^{-(1-\alpha)^2/8\alpha}$. This slower  decay is due to the presence of long-range algebraically localized eigenstates, centered beyond $x=L$, whose overlap with the initial wave function is significant, thus enhancing the probability of finding the particle at $x>L$.
Note that this effect is expected to be less important when the eigenstates are more strongly localized, \ie, when $\alpha$ vanishes. This is consistent with the equality of the two exponents, $(1-\alpha)^2/8\alpha \simeq 1/8\alpha$ in the limit $\alpha\to 0$.

\section{Experimental observation}
The transmission can be measured directly via the Landauer conductance~\cite{landauer_spatial_1957,buttiker_symmetry_1988,akkermans_mesoscopic_2007} in mesoscopic materials~\cite{depicciotto2001} or ultracold atoms~\cite{brantut_conduction_2012,stadler_observing_2012,krinner_superfluidity_2013}, for instance.
While the discussion below is generic, we focus on ultracold atoms for the sake of concreteness. From a practical point of view, these systems offer key features, such as the possibility to control the disorder and to accommodate arbitrary large forces without damage.
The Landauer conductance is defined as the ratio of the current $I$ induced by the potential imbalance between two charge reservoirs to their potential difference $\Delta V$ 
(see Fig.~\ref{fig:reservoirs_deltaV})~\cite{buttiker_symmetry_1988}.
In the simplest case where $\Delta V$ is measured inside the reservoirs (two-terminal scheme), it reads
$G_2(L) = \frac{S}{2\pi\hbar\Delta V}\int_{\mu-\Delta V/2}^{\mu+\Delta V/2} dE~T(E,L)$,
where $\mu \pm \Delta V /2$ are the chemical potentials of the two reservoirs and $S$ is the spin degeneracy.
The various longitudinal energy levels give statistically independent contributions since they probe different Fourier components of the disorder, which are independent. Assuming that the potential difference $\Delta V$ is large compared to the longitudinal energy level spacing inside the guide but smaller than the energy variation scale of $T(E,L)$, the finite potential difference $\Delta V$ may realize disorder averaging~\cite{mailly_reduction_1989,mailly_sensitivity_1992} and $G_2(L) \simeq \overline{T}(\mu,L)$.
Moreover, in real experiments, the guide  is not purely one dimensional and may contain tens of transverse modes, which realize further effective disorder averaging.
For short guides, however, one might rather measure the typical conductance
$T_\textrm{\tiny typ} \sim \exp\left[\overline{\ln(T)}\right]$.

Realistic models of disorder should include finite-range correlations.
The metrics $s(L)$ then reads
\begin{equation}
s(L)
=\int_0^L dx\, \frac{m\tilde{C}[2k(x)]}{2\hbar^2(E+Fx)},
\end{equation}
which increases as 
$s(L)\sim \int_{k(0)}^{k(L)} dk\, \tilde{C}(2k)/k$.
Since any model of disorder with finite-range correlations has an integrable two-point correlation function $\tilde{C}(k)$ in reciprocal space, the metrics $s(L)$  saturates to a finite value when $L\to\infty$ in the presence of finite-range correlations. Hence, the distribution $P(T,L)$, and all disorder-average functions of $T$ saturate to a nonzero value when $L \rightarrow \infty$. This effect suppresses algebraic localization and entails that any finite-range correlations induce delocalization.
Similarly, an increase of the bias force also entails delocalization. For instance, for white-noise disorder and $F(x) \sim x^{\varepsilon}$ with $\varepsilon > 0$, we find $s(L) \sim \textrm{const} - 1/{L}^{\varepsilon}$, which also saturates when $L \rightarrow \infty$.
Delocalization, as signaled by the saturation of
$\overline{T(L)}$ and $\overline{\ln T(L)}$, is confirmed by numerical calculations
for
either Gaussian correlations, $C(x)=\frac{\Ur}{\sigmar\sqrt{2\pi}}\exp\left(-x^2/2\sigmar^2\right)$, and a uniform bias force [see Fig.~\ref{fig:graph_universality}(b), magenta circles]
or white-noise disorder and a slightly linearly increasing force $F(x)=F+m\omega^2x$ [see Fig.~\ref{fig:graph_universality}(b), brown diamonds].

Delocalization appears for any realistic model of disorder with finite-range correlations.
Assuming that $C(x)$ decays on the length scale $\sigmar$, the transmission crosses over from algebraic decay to saturation for a length $L^*$ given by $\Delta k(L^*)\sigmar \sim 1$ with $\Delta k(L)=k(L)-k(0)$. In the two-Fermi terminal configuration of Fig.~\ref{fig:reservoirs_deltaV},
$k(0)=k_{\textrm{\tiny F}}=\sqrt{2mk_{\textrm{\tiny B}}\theta_{\textrm{\tiny F}}}/\hbar$
is the Fermi wave vector of the left-hand-side reservoir,
with $\theta_{\textrm{\tiny F}}$ the Fermi temperature and $k_{\textrm{\tiny B}}$ the Boltzmann constant.
Using $\Delta k(L) \sim \partial_L k(0)\times L \sim m F L / \hbar^2 k_{\textrm{\tiny F}}$,
we find $L^* \sim \sqrt{\frac{2 k_{\textrm{\tiny B}}\theta_{\textrm{\tiny F}}}{m \sigmar^2}}\frac{\hbar}{F}$.
For mesoscopic channels designed in ultracold atomic systems~\cite{brantut_conduction_2012,krinner_superfluidity_2013}, where typically
$\theta_{\textrm{\tiny F}} \sim 500$~nK and $\sigmar \sim 0.5~\mu$m,
and assuming that the force results from gravity on $^6$Li atoms, $F \simeq 10^{-25}$~N,
we find $L^* \sim 80~\mu$m, which is of the order of magnitude of the experimentally realizable channel lengths. Note that the effective force can be reduced by compensating gravity with a magnetic levitation field~\cite{jendrzejewski_three-dimensional_2012} or using a nonvertical geometry.
Whatever their strength, these delocalization effects can, however, be incorporated into the metrics to test the universal relation~(\ref{eq:DistrAbrikosov}).
Indeed, as shown on Fig.~\ref{fig:graph_universality}(a),
we recover a universal behavior of the transmission by rescaling the Euclidean distance $L$ to the metrics $s(L)$
for correlated disorder, as well as nonuniform bias field~\footnote{We have checked that the distribution of transmission, Eq.~(\ref{eq:DistrAbrikosov}), still holds in the presence of finite correlation lengths and nonuniform bias force [see the inset of Fig.~\ref{fig:graph_universality}(b)].}.

Another matter of concern is finite temperature, which is typically a fraction of the Fermi temperature in ultracold atoms.
Using the Sommerfeld expansion of the current-potential characteristic function, we find
\begin{equation}
\overline{G_2}(L)=G_0\left\{\overline{T}[s(L)]+\frac{\pi^2 (k_\textrm{\tiny B}\theta)^2}{6}A(L)\right\},
\end{equation}
where, for a constant force $F$,
\begin{eqnarray}
A(L) & = &
\frac{1}{F} \frac{\partial}{\partial \mu}\left[\frac{1}{\ell_-(L)}-\frac{1}{\ell_-(0)}\right]  \frac{\partial\overline{T}}{\partial s}
\label{eq:A} \\
& & +\frac{1}{F^2}\left[\frac{1}{\ell_-(L)}-\frac{1}{\ell_-(0)}\right]^2  \frac{\partial^2\overline{T}}{\partial s^2}.
\nonumber
\end{eqnarray}
For the parameters above and the typical disorder strength $\Vr/k_{\textrm{\tiny B}}\sim 0.5~\mu$K,
we find that finite-temperature effects contribute the conductance $\overline{G_2}$ from less than $2\%$ for $\theta\sim0.1\,\theta_\textrm{\tiny F}$ up to $15\%$ for $\theta\sim0.3\,\theta_\textrm{\tiny F}$.
Since the quantity $A(L)$ is not a universal function of the rescaled length $s(L)$, such finite-temperature effects break universal scaling. However, if the system is sufficiently large that $\ell_-(L) \gg \ell_-(0)$, 
then $\overline{G_2}(L)$ becomes a pure function of $s(L)$ and
universal scaling is recovered.

\section{Conclusion and outlook}
In summary, we have computed the statistical distribution of transmission in a disordered matter waveguide in the presence of a bias field. For white-noise disorder and a uniform bias force, we have shown analytically that the transmission decays algebraically, irrespective of the value of the force,
in agreement with numerical calculations~\cite{soukoulis_localization_1983}. This behavior differs from what is predicted in the expansion of a wave packet, which features a delocalization transition~\cite{prigodin_one-dimensional_1980} (see also Refs.~\cite{delyon_power-localized_1984,bentosela1985}).
This striking difference has been traced back to the long-range character of the algebraic decay of the localized eigenstates in the presence of the bias field. We have proposed a concrete observation of the transmission decay using ultracold atoms.
While finite-range disorder correlations or nonuniform bias force entails systematic delocalization, we have shown that a universal behavior can be recovered using appropriate rescaling. We have also found that finite-temperature effects can be ignored for a long-enough waveguide.

Ultracold atoms also offer an interesting alternative to measure the transmission. The idea is to create a trapped Bose-Einstein condensate, which is less sensitive to finite temperatures than fermions, above a finite-size disordered region. Releasing the condensate from the trap as done in Refs.~\cite{billy_direct_2008,roati_anderson_2008,kondov_three-dimensional_2011,jendrzejewski_three-dimensional_2012}, for instance, the atoms fall down in the gravity field, possibly partially compensated by a levitation magnetic field to control the force. The number of transmitted atoms is then proportional to the average transmission. The same experiment with the condensate created directly in an infinite-size disorder will in turn yield the asymptotic average density. This offers a single platform to compare directly the two quantities $\overline{T(x)}$ and $\overline{n(x)}$.
More precisely, this scheme would provide the quantities $\overline{T}$ and $\overline{n}$ integrated over the energy distribution of the falling condensate~\cite{sanchez-palencia_anderson_2007,piraud2011}. This issue can, however, be circumvented by using an energy-selective radio-frequency transfer of atoms from an atomic state insensitive to the disorder to another state sensitive to the disorder~\cite{pezze2011a}.

\section{Acknowledgments}
We thank Boris Altshuler, Gilles Montambaux, Marie Piraud, and Christian Trefzger for useful discussions.
This research was supported by the
European Commission FET-Proactive QUIC (H2020 Grant No. 641122).
It was performed using HPC resources from GENCI-CCRT/CINES (Grant No. c2015056853).
Use of the computing facility cluster GMPCS of the LUMAT federation (FR LUMAT 2764) is also acknowledged. 

\bibliography{paper1}
  
  
 \renewcommand{\theequation}{S\arabic{equation}}
 \setcounter{equation}{0}
 \renewcommand{\thefigure}{S\arabic{figure}}
 \setcounter{figure}{0}
 \onecolumngrid  
     
 
 \newpage

 {\center \bf \large --Supplemental Material-- \\}
 {\center \bf \large Effect of a bias field on disordered waveguides: Universal scaling of conductance and application to ultracold atoms \\ \vspace*{1.cm}
 }

 In this supplemental material, we provide details about the transfer matrix formalism for an inhomogeneous medium and the derivation of the Fokker-Planck equation [Eq.~(\ref{eq:FPE}) of the main paper].
 
 \bigskip
 \section{Transmission and reflection coefficients}
 To compute the transmission coefficient of a particle submitted to a bias force $F(x)$ through a disordered sample in the space interval $[0,L]$,
it is convenient to define the semiclassical kinetic energy $K(x) \equiv E+\int_0^x dx'\, F(x')$ and the associated wave vector $k(x)=\sqrt{2mK(x)}/\hbar$.
 At any position $x$, we may write the particle wave function $\psi(x)$ and its derivative $\partial_x\psi(x)$ in the form
 \begin{equation}
 \begin{pmatrix}
 \psi(x)\\ 
 \partial_x \psi(x)
 \end{pmatrix}
 =
 \begin{pmatrix}
 e^{ik(x)x} & \e^{-ik(x)x}\\ 
 ik(x)e^{ik(x)x} & -ik(x)e^{-ik(x)x}
 \end{pmatrix}
 \begin{pmatrix}
 \psi_+(x)\\ 
 \psi_-(x)
 \end{pmatrix}
 \end{equation}
 A unique solution $\left(\psi_+(x),\psi_-(x)\right)$  exists provided the determinant of the above matrix does not vanish, \ie, $k(x)\ne 0$.
 
 The particle flux, $j(x) \equiv \frac{\hbar}{2im} \left(\psi^*\partial_x\psi - \psi\partial_x\psi^*\right)$ then reads $j(x)=j_+(x)+j_-(x)$, where
 \begin{equation}
 j_{\pm}(x)=\pm \frac{\hbar k(x)}{m}\left|\psi_\pm (x)\right|^2
 \end{equation}
 are the right-moving ($+$) and left-moving ($-$) fluxes.
 The transmission coefficient is then defined as the ratio of right-moving fluxes at the boundaries of the sample in the case where the incident flux is right-moving, \ie, $j_-(L)=0$, and reads
 \begin{equation}
 T(L)
 \equiv \frac{j_+(L)}{j_+(0)}
 = \frac{k(L)}{k(0)}\frac{\left|\psi_+(L)\right|^2}{\left|\psi_+(0)\right|^2}.
 \end{equation}
 Under the same assumption of right-moving incident flux,
 the reflection coefficient is defined as the ratio of left-moving emergent flux and the right-moving incident flux, and reads
 \begin{equation}
 R(L) \equiv \frac{\vert j_-(0)\vert}{j_+(0)}
 = \frac{\left|\psi_-(0)\right|^2}{\left|\psi_+(0)\right|^2}.
 \end{equation}
 
 \bigskip
 \section{Scattering matrix}
 Consider now a finite sample in the interval $[x_1,x_2]$, where $0 \leq x_1<x_2 \leq L$.
 We define the scattering matrix
 \begin{equation}
 \bold{S}=
 	\begin{pmatrix}
 		r & t'\\
 		t& r'
 	\end{pmatrix}
 \end{equation}
 such that
 \begin{equation}
 \begin{pmatrix}
  	\sqrt{k(x_1)}\psi_-(x_1)\\
  	\sqrt{k(x_2)}\psi_+(x_2)
  \end{pmatrix}
  =
  	\bold{S}
  	\begin{pmatrix}
  	\sqrt{k(x_1)}\psi_+(x_1)\\
  	\sqrt{k(x_2)}\psi_-(x_2)
  	\end{pmatrix}.
 \end{equation}
 Particle-flux conservation between $x_1$ and $x_2$ imposes that the scattering matrix is unitary,
 $\bold{S}^\dagger\bold{S}=\mathbb{1}$. Moreover, time-reversal symmetry entails $\bold{S}^*=\bold{S}^{\dagger}$. Those two relations lead to the usual relations
 \begin{equation}
 \left\{
 \begin{aligned}
 & t=t' \\
 & | r |^2+ | t |^2 = | r' |^2 + | t |^2=1 \\
 & t^*r'+r^*t=0.
 \end{aligned}
 \right.
 \end{equation}
 Straightforward calculations then lead to the usual relations for the transmission and reflection coefficients of samples between the points $x_1$ and $x_2$,
 \begin{equation}
 T=|t|^2
 \qquad \textrm{and}\qquad
 R=|r|^2=|r'|^2.
 \end{equation}

 \bigskip
 \section{Transfer matrix}
 We now define the transfer matrix $\bold{T}(x_2,x_1)$ such that
 \begin{equation}
 \begin{pmatrix}
  	\sqrt{k(x_2)}\psi_+(x_2)\\
  	\sqrt{k(x_2)}\psi_-(x_2)
  \end{pmatrix}
  =
  	\bold{T}(x_2,x_1)
  	\begin{pmatrix}
  	\sqrt{k(x_1)}\psi_+(x_1)\\
  	\sqrt{k(x_1)}\psi_-(x_1)
  	\end{pmatrix}.
 \end{equation}
 Straightforward calculations yield
 \begin{equation}
 \bold{T}=
 	\begin{pmatrix}
 		1/t^* & r'/t\\
 		-r/t & 1/t
 	\end{pmatrix}
 \end{equation}
 
 Transfer matrices can then be chained, \ie,
 \begin{equation}
 \bold{T}(x_n,x_1)=\bold{T}(x_{n},x_{n-1})\bold{T}(x_{n-1},x_{n-2}) ... \bold{T}(x_2,x_1).
 \end{equation}
 
  \bigskip
 \section{Fokker-Planck equation}
 Considering two samples in the intervals $[0,x]$ and $[x,x+\Delta x]$ respectively, where the sample [0,x] has transmission coefficient $T(x)$ and reflection coefficient $R(x)=1-T(x)$, and
 the sample $[x,x+\Delta x]$ has transmission coefficient $T_{\Delta x}(x)$ and reflection coefficient $R_{\Delta x}(x)$, the transmission coefficient of the sample $[0,x+\Delta x]$ can be calculated from the product of the two transfer matrices. It yields
 \begin{equation}
 T(x+\Delta x)=\frac{T(x)T_{\Delta x}(x)}{|1-\sqrt{R(x) R_{\Delta x}(x)}\e^{i\theta_{\Delta x}(x)}|^2},
 \end{equation}
 where $\theta_{\Delta x}(x)$ is the phase accumulated during one total internal reflection at point $x$, with $r'(x)r_{\Delta x}(x)=|r'(x)r_{\Delta x}(x)|\e^{i\theta_{\Delta x}(x)}$.

 Let us define the backscattering mean free path at kinetic energy $K(x)$, 
\begin{equation}\label{eq:MFpath}
\ell_-(x) \simeq {2\hbar^2K(x)}/{m\tilde{C}[2k(x)]},
\end{equation}
where $\tilde{C}[2k(x)]$ is the disorder power spectrum.
For weak disorder, \ie, $\ell_-(x)\gg \lambda(x),\sigmar$, we may choose intermediate elementary lengths $\Delta x$, such that $\lambda (x),\sigmar \ll \Delta x \ll \ell_-(x)$, where $\lambda (x)=2\pi/k(x)$. 
Since $\Delta x \ll \ell_-(x)$, the nonvanishing value of the reflection coefficient $R_{\Delta x}(x)$ results from typically less that one scattering and it may thus be computed in the single-scattering approximation, provided the local mean free path $\ell_-(x)$ is well defined on the elementary cell $\Delta x$. This implies that $\partial_x \ell_-(x) \Delta x\ll \ell_-(x)$, or equivalently it corresponds to assuming that the work of the force on the length of the elementary cell is small,
$F(x)\Delta x \ll K(x),
\hbar^2 k(x)/2m\partial_k\ln\tilde{C}[2k(x)]$. 
We then find
${\overline{R_{\Delta x}(x)} \simeq \Delta x/\ell_-(x)} \ll 1$, where the overline denotes averaging over the disorder. Consequently, $R_{\Delta x}(x)\ll 1$ and we may use the following expansion of $\Delta T (x)\equiv T(x+\Delta x) - T(x)$:
 \begin{equation}
 \Delta T (x)=T(x)\left\{ 2\sqrt{(1-T(x))R_{\Delta x}(x)}\cos\theta_{\Delta x}(x) + R_{\Delta x}(x) \left[T(x)-2+4(1-T(x))\cos^2\theta_{\Delta x}(x)\right]\right\} +O(R_{\Delta x}(x) ^{3/2})  
  \end{equation} 
 The transmission coefficient is thus governed by a stochastic process when the system length $x$ increases. The Kramers-Moyal expansion of the corresponding master equation for the probability distribution of the transmission coefficient at a given length, $P(T,x)$, reads
 \begin{equation}{\label{eq:KME}}
 \frac{\partial P(T,x)}{\partial x}=\sum_{n=1}^{+\infty}\frac{(-1)^n}{n!}\frac{\partial^n}{\partial T^n}\left[M_n(T) P(T,x)\right]
 \end{equation}
 with
 \begin{equation}
 M_n(T)=\left . \frac{\overline{(\Delta T(x))^n}}{\Delta x}\right|_{\Delta x\to 0}.
 \end{equation}
 Assuming that the quantity $\theta_{\Delta x}(x)$ is uniformly distributed on $2\pi$ and knowing $\overline{R_{\Delta x}(x)}$, 
 the average on both quantities can be performed independently. We then find
 \begin{equation}
 \begin{aligned}
 M_1=-\frac{T^2(x)}{\ell_-(x)}\hspace*{2cm}&
 M_2=\frac{2T^2(x)(1-T(x))}{\ell_-(x)}\hspace*{2cm}&
 M_n=0\text{ for }n\ge 3.
 \end{aligned}
 \end{equation}
 The Kramers-Moyal expansion~\eqref{eq:KME} thus reduces to its first two moments, which yields the Fokker-Planck equation~\eqref{eq:FPE} of the main paper.
 
\end{document}